\documentclass[twocolumn,aps,prl,superscriptaddress,showpacs]{revtex4}
\usepackage{graphicx}
\usepackage{amsmath}
\usepackage{amssymb}

\usepackage[latin1]{inputenc}
\usepackage[T1]{fontenc}
\usepackage{ae,aecompl}

\usepackage[colorlinks=true,urlcolor=blue,citecolor=blue,hyperfootnotes=false]{hyperref}

\newcommand{\betas}{{\beta_{\text{s}}}} % spin transfert torque parameter, to be distinguished from the depinning exponent
\newcommand{\bOmega}{\mathbf{\Omega}}
\newcommand{\cW}{\mathcal{W}}         % winding number
\newcommand{\bx}{\mathbf{x}}
\newcommand{\fc}{f_{\text{c}}}
\newcommand{\vs}{v_{\text{s}}}
\newcommand{\js}{j_{\text{s}}}
\newcommand{\fw}{f_{\text{w}}}
\newcommand{\vw}{v_{\text{w}}}
\newcommand{\Hext}{f_\textnormal{ext}}
\renewcommand{\epsilon}{\varepsilon}
\renewcommand{\phi}{\varphi}

\begin{document}

\title{Depinning of domain walls with an internal degree of freedom} 

\author{V. Lecomte}
\affiliation{DPMC-MaNEP, University of Geneva, 24 quai Ernest-Ansermet, 1211 Geneva 4, Switzerland}

\author{S. E. Barnes}
\affiliation{DPMC-MaNEP, University of Geneva, 24 quai Ernest-Ansermet, 1211 Geneva 4, Switzerland}
\affiliation{Physics Department, University of Miami, Coral Gables, FL 33124, USA}

\author{J.-P. Eckmann}
\affiliation{Department of Theoretical Physics and Section de Math\'ematiques, University of Geneva, 24 quai Ernest-Ansermet, 1211 Geneva 4, Switzerland}

\author{T. Giamarchi}
\affiliation{DPMC-MaNEP, University of Geneva, 24 quai Ernest-Ansermet, 1211 Geneva 4, Switzerland}

\begin{abstract}
  Taking into account the coupling between the position of the wall
  and an internal degree of freedom, namely its phase $\phi$, we
  examine, in the rigid wall approximation, the dynamics of a magnetic
  domain wall subject to a weak pinning potential.
  We determine the corresponding force-velocity characteristics, which display
  several unusual features when compared to standard depinning laws.
  At zero temperature, there exists a bistable regime for low forces,
  with a logarithmic behavior close to the transition.  For weak
  pinning, there occurs a succession of bistable transitions
  corresponding to different topological modes of the phase evolution.
  At finite temperature, the force-velocity characteristics  become
  non-monotonous. We compare our results to recent experiments on
  permalloy nanowires.
  
\end{abstract}

\maketitle

%%%%%%%%%%%%%%%%%%%%%%%%%%%%%%%%%%%%%%%%%%%%%%%%%%%%%%%%%%%%%%%%%%%%%%

Many physical systems comprise different phases which coexist and are
separated by an interface. Examples range from
magnetic~\cite{lemerle_domainwall_creep,bauer_ferre_dw-dipolar_prl2005,yamanouchi_dw_prl1996,metaxas_ferre_prl2007}
or ferroelectric~\cite{paruch_PZT_prl2005,paruch_triscone_PZT_apl2006}
domain walls (DWs), to growth
surfaces~\cite{barabasi_stanley_book,krim_growth_review} or contact
lines~\cite{moulinet_distribution_width_contact_line}.
Common to this large variety of phenomena is a macroscopic description
within which the interface properties are well described by a
competition between the elasticity, which tends to minimize the
interface length, and the local potential, whose valleys and hills
deform the interface so as to minimize its total energy.

Such interfaces are described by the theory of disordered elastic
systems~\cite{kardar_phys-rep1998,giamarchi_book_young},
which explains well their static (e.g., roughness at equilibrium,
correlation functions) as well as dynamical features (transient
regime, response to a field).
The existence of a threshold force $\fc$ below which the system is
pinned is a crucial feature of the zero-temperature motion of such an
interface.
When $f\gtrsim \fc$ the velocity $v\sim(f-\fc)^\beta$ is characterized
by a depinning exponent $\beta$, while at finite temperature $v\sim
T^\psi$ at $f=\fc$ defines the thermal exponent $\psi$.
Some predicted exponents are in very good agreement with measurements,
e.g., in magnetic~\cite{lemerle_domainwall_creep} or
ferroelectric~\cite{paruch_triscone_PZT_apl2006} films,
while discrepancies remain for contact
lines~\cite{moulinet_distribution_width_contact_line} or for magnetic
wires~\cite{Yamanouchi-SCI-2007}, and one can ask what are the missing
ingredients in the description.
In particular, it is usual in the macroscopic description to specify
only the position of the interface, discarding, a priori, as
irrelevant internal structures. Here we investigate how this position
couples to an internal degrees of freedom, and how this coupling is
manifested in experiment.

\begin{figure}[t] %%%% FIGURE NUMBER 1
  \centering
  \includegraphics[width=.93\columnwidth]{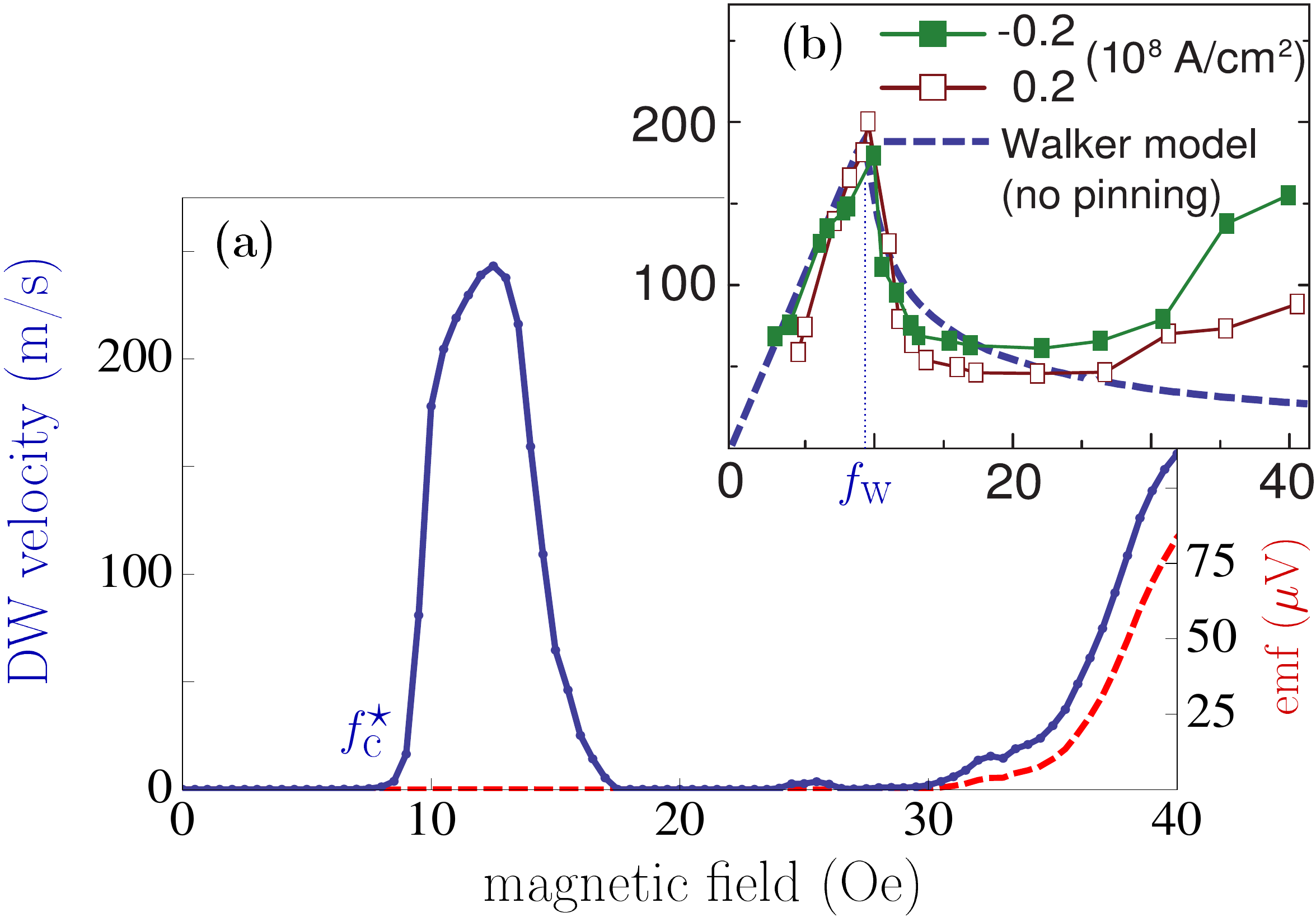}
  \vspace*{-2mm}
  \caption{(color online) \textbf{(a)} Finite temperature, zero
    current ($\js=0$) force-velocity  curves (blue, solid) and
    emf $\frac{\hbar}{e}\langle\dot\phi\rangle$ (red, dashed) for a DW
    with pinning. Parameters correspond to experiments shown in
    \textbf{(b)}. Adapted from Ref.~\cite{Parkin-SCI-2008}, the data
    (squares) are for small but opposite currents. In the absence of
    pinning (Walker model, dashed blue line) there is no second rapid
    increase in $v$ manifest in the data.}
  \vspace*{-1mm}
  \label{fig:comparison_experiment}
\end{figure}

In magnetic systems, the DW position is generically coupled to an
internal degree of freedom (a spin phase $\phi$ \footnote{Although
  this coupling is well known in the magnetic DW
  community~\cite{malozemoff_Slonczewski_DW}, it has to our knowledge
  always been discarded in interface physics.}).
An interesting case is the motion of a $180^\circ$\:DW in a narrow
ferromagnetic thin film, which has been the subject of intense
experimental study because of its importance for
spintronics~\cite{yamanouchi_dw_prl2004,parkin-nature-2006,Yamanouchi-SCI-2007,Parkin-SCI-2008}.
It is known~\cite{schryer-walker_jap-1974}, in the absence of pinning,
that $v(f)$ increases up to a characteristic `Walker' force $\fw$
above which the velocity actually {\it decreases\/} up to values $f\gg
\fw$\:(Fig.\ref{fig:comparison_experiment}).
In contrast standard interface theory~\cite{kardar_phys-rep1998} takes
pinning into account but not the phase, yielding a monotonous $v(f)$.
An approach combining these two ingredients is missing.

In this Letter, we develop such an approach and show there are
dramatic changes as compared with both the Walker picture and standard
interface theory. Specifically,
the DW is pinned up to a force $\fc^\star$ above which the depinning
law is bistable and logarithmic.
Even more strikingly, as $f$ is increased further, the velocity falls
back to zero until a second depinning transition
occurs\:(Fig.\ref{fig:comparison_experiment}).  This is followed by a
cascade of such transitions until finally $v(f)$ becomes monotonous.
Upon adding the effects of a finite temperature, this offers a natural
explanation of the two peaks of $v(f)$ observed in
experiments~\cite{Parkin-SCI-2008}.

%%%
%%% BULK MODEL, NOTATIONS
%%%

We consider an uniaxially anisotropic ferromagnetic medium with
position-dependent magnetization of direction
$\bOmega=(\sin\theta\cos\phi,\sin\theta\sin\phi,\cos\theta)$
with easy $z$-axis and hard $y$-axis with respective anisotropy
constants $K$ and $K_\perp$.  With spin stiffness $J$, the energy
$E[\bOmega]$ is~\cite{landau-electromag-media-english}
\begin{align*}
  E=  \frac 12 \int\! \text{d}^3x \Big\{
    J \Big[(\nabla\theta)^2&+\sin^2\theta(\nabla\phi)^2\Big] \\[-2mm] & + 
    K\,\sin^2\theta +
    K_\perp\sin^2\theta\cos^2\phi
  \Big\}
\end{align*}
The corresponding Landau-Lifshitz dynamics reads
\begin{equation}
  \big(\partial_t + \vs \nabla \big) \bOmega 
= 
  \bOmega\times \big(H+\mathbf{\tilde\eta}\big)
  -\bOmega \times 
  \big(\alpha\partial_t + \betas \vs  \nabla\big)\bOmega
\label{eq:bulk_eq-motion}
\end{equation}
Here, $H=-\delta E/\delta\bOmega+\Hext$ and $\Hext$ is the external
field and $\alpha$ is the Gilbert damping, which accounts for
dissipation.  The velocity $\vs$\:\footnote{ $\vs=P \js /(e\rho_{\text
    s})$ with $P$ the current polarization, $e$ the carrier charge and
  $\rho_{\text s}$ the spin density.} is proportional to the
spin-polarized current density $\js$ and $\betas$ is the
current-induced relaxation.  The white noise $\mathbf{\tilde\eta}$
accounts for thermal fluctuations for temperature $T$\:\footnote{It
  has zero mean and variance~\cite{brown_FDR-gilbert_PR1963}: 
  $ \big\langle\tilde\eta_i(\bx,t)\tilde\eta_j(\bx',t')\big\rangle =
  2\hbar^{-1}\alpha k_B T
  \,\delta(\bx'-\bx)\,\delta(t'-t)\,\delta_{ij} $ .}.
Below the Walker field $\fw=\frac 12 \alpha K_\perp$, and for $T=0$,
there exists a solution to~\eqref{eq:bulk_eq-motion} for constant
$\Hext$ \:\cite{schryer-walker_jap-1974}: $\theta(x,t)=2\arctan
\exp\big[(x/\lambda-\vw t)/\xi \big]$ with
$\xi=[1+(K_\perp/K)\:\sin\phi]^{-1/2}$ and constant
$\phi(x,t)=\frac 12\arcsin\Hext/\fw$.
This represents a N\'eel DW of width $\lambda=\sqrt{J/K}$ and
velocity $\vw=\xi\Hext/\fw$\:\footnote{Bloch DWs are treated similarly~\cite{tatara-kohno_JPSJpn2008}.}.

%%%
%%% ANSATZ, EFFECTIVE EQUATIONS OF MOTION
%%%

In more general situations, this domain wall solution with $\vw t$
replaced by the actual position $r(t)$ and with this and $\phi(t)$
considered as parameters is used as an ansatz. In the rigid wall
approximation (constant $\xi$), one obtains the effective
equations~\cite{schryer-walker_jap-1974,malozemoff_Slonczewski_DW,slonczewski_jmmm1996,tatara-kohno_DWcurrent_prl2004,barnes_maekawa_prl2005,duine_thermalDW_prl2007}
\begin{align}
  {\alpha}\partial_t r -\partial_t\phi  -{\betas} \vs 
  &=
  \Hext(r) \:+\:\eta_1
  \label{eq:eq_motion1}
\\
  \alpha \partial_t \phi +  \partial_tr -  \vs 
  &=
  - \tfrac 12 K_\perp \sin 2\phi \:+\:\eta_2
  \label{eq:eq_motion2}
\end{align}
($\lambda=1$ by choice of length units\:\footnote{We also translated
  $\phi\to\phi+\frac\pi 2$ for convenience.}.)  We split the external
field $\Hext(r)=f-\:V'(r)$ into a constant `depinning' (or `tilt')
force $f$ and a `pinning' force $-V'(r)$ deriving from a potential
$V$. The effective thermal noise is now~\cite{duine_thermalDW_prl2007}
$\langle\eta_i(t)\eta_j(t')\rangle = 2(\hbar N)^{-1}\alpha k_B T
\delta(t'-t)\,\delta_{ij} $
where $N=2\lambda A/a^3$ is the number of spins in the DW, of section
$A$.
For constant fields $f>\fw$, this ansatz reproduces very
accurately~\cite{schryer-walker_jap-1974} the numerical solution of
the bulk equation~\eqref{eq:bulk_eq-motion}.
We checked that this result also extends to the case of a non-uniform
potential landscape~\cite{non-monotonous_long_in-prep}.

Having simplified~\eqref{eq:bulk_eq-motion} to
(\ref{eq:eq_motion1}-\ref{eq:eq_motion2}), we further restrict our
study to the case $\js=0$: current effects will be considered
elsewhere~\cite{non-monotonous_long_in-prep}.
The potential $V(r)$ should reflect the pinning effects of impurities
or local variations in the couplings.  A proper treatment of a
realistic disordered $V(r)$ is a delicate task and in order to gain
insights into the full problem we take a periodic
$V(r)= \frac{1}{\kappa}\sin \kappa r$\:\footnote{For an overdamped particle in an arbitrary
  potential~\cite{derrida_hopping-particles_JSP1982,ledoussal_anomalous_diffusion},
  taking $V(r)$ random yields $r(t)\sim t^\gamma$ and $v(f)$ is not
  finite in
  general,
  while taking $V(r)$ periodic leads to standard depinning with
  $\beta=\frac 12$ and $\psi= \frac 13$.}.

%%%
%%%  SIMPLE CONSIDERATIONS
%%%

We first consider the zero-temperature motion.
Before embarking into a thorough analysis
of~(\ref{eq:eq_motion1}-\ref{eq:eq_motion2}), let us gain some
insights from simple considerations.
At $f=0$, the wall is pinned in one of the minima of
$V(r)=\frac{1}{\kappa}\sin\kappa r$.
%
% We thus anticipate to have a threshold force beyond which
% the soliton moves at non-zero mean velocity in the large time
% limit.
%
There exists a characteristic force $\fc$ beyond which local minima of
the tilted potential $\frac{1}{\kappa}\sin \kappa r -fr$ disappear
(here $\fc=1$).
Ignoring the variable $\phi$, the wall would start to move at $f=\fc$
and acquire a finite mean velocity at long times for $f>\fc$, because
of damping.
But since $r$ is coupled to $\phi$, the wall may store enough kinetic
energy in $\phi$ to cross barriers for forces less than $\fc$, hence
shifting the depinning transition to some $\fc^\star<\fc$.
For $f$ between these values, the system is bistable: depending on the
initial condition, the wall is either pinned in a minimum or slides
down in the tilted landscape while $\phi$ oscillates around its own
minimum.
Moreover, the periodicity of $\phi$ can induce an unexpected effect:
%
% BE PREPARED FOR THE UNEXEPECTED
%
increasing $f$ makes $\phi$ cross its own barrier and fall into its
next minimum, but this has a cost: dissipation increases and $\phi$
cannot give back enough kinetic energy to $r$. This intuitive picture
explains the valley appearing in $v(f)$
(Fig.\ref{fig:comparison_experiment}) until the depinning force
injects enough energy to reach another regime where both $\phi$ and
$r$ increase in time.

\begin{figure}[t] %%%% FIGURE NUMBER 2
  \centering
  \includegraphics[width=.9\columnwidth]{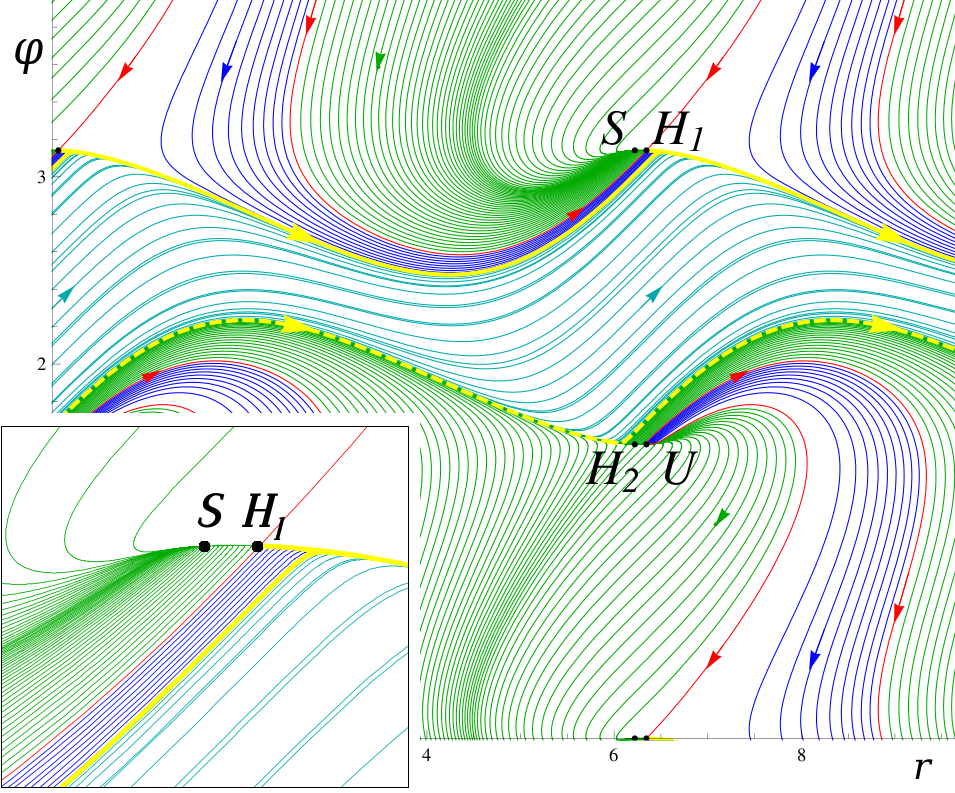}
  \vspace*{-1.5mm}
  \caption{
    (color online) Phase space trajectories $(r,\phi)$ for $f$ in the bistable
    regime ($\fc^\star<f<\fc$). $S$, $U$, $H_{1,2}$ are the fixed
    points. Blue and turquoise trajectories converge
     to the attracting
    limit cycle (yellow).  Those in green end at the stable fixed
    point $S$.  Separatrices (red) mark the boundaries between the
    corresponding regions.  The repulsive limit cycle (dashed yellow)
    is also a separatrix. }
  \vspace*{-4mm}
  \label{fig:phase_space}
\end{figure}

%%%
%%%  PHASE SPACE ANALYSIS
%%%

The analysis of~(\ref{eq:eq_motion1}-\ref{eq:eq_motion2}) can be put
on a firm basis by considering the phase space of $(r,\phi)$, which is
a torus of period $2\pi/\kappa$ in $r$ and $\pi$ in $\phi$.
We determine the nature of the possible trajectories. These cannot
cross in phase space, but can meet at the fixed points (steady
solutions of~(\ref{eq:eq_motion1}-\ref{eq:eq_motion2})).
Trajectories approaching a stable fixed point have zero mean velocity
$v\equiv\langle\partial_t r\rangle$, 
unlike those moving along a limit cycle.
In the latter case the particle covers one spatial period
$2\pi/\kappa$ over a period of time $\tau$, so that the average
velocity is given by $v=2\pi/(\kappa\tau)$.
We will thus determine $v(f)$ through $\tau$.

\begin{figure}[t] %%%% FIGURE NUMBER 3
  \centering
  \includegraphics[width=.9\columnwidth]{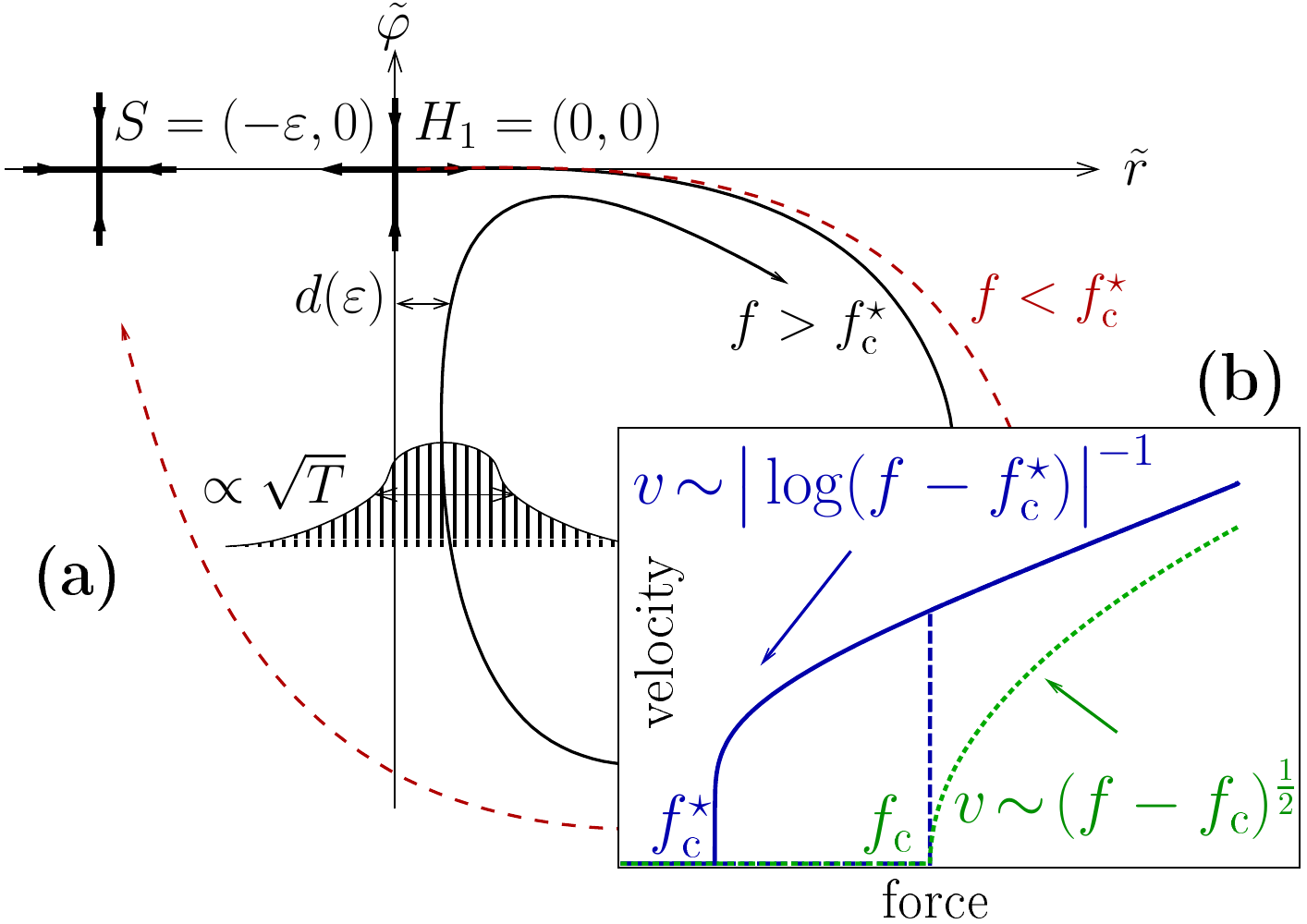}

  \vspace*{-1mm}
  \caption{(color online) 
  \textbf{(a)}
   Sketch of the phase space for the coordinates $(\tilde r,\tilde\phi)$,
   and  homoclinic bifurcation. For $f<\fc^\star$ (red, dashed),
   trajectories all end at $S$. For $f>\fc^\star$ (black, solid), the 
   line starting from the unstable direction of $H_1$ converges to the 
   limit cycle.
   At $T>0$, the random evolution has  Gaussian probability distribution of
   variance $\propto\sqrt{T}$.  
   \textbf{(b)} The two types of depinning laws at $T=0$.  For
   large $K_\perp$ (green, dotted) depinning occurs with $\beta=\frac
   12$.  For intermediate $K_\perp$ (blue, solid) there is a bistable
   regime of force ($\fc^\star<f<\fc$) with a zero-velocity branch
   and a non-zero one in $1/\log$.}
  \vspace*{-2mm}
  \label{fig:homoclinic-scheme}
\end{figure}

We remark that (\ref{eq:eq_motion1}-\ref{eq:eq_motion2}) has no fixed
points for $f>\fc$, which means that the DW moves with non-zero
velocity.
For $f<\fc$, there are four fixed points of coordinates
(Fig.\ref{fig:phase_space}): $H_1=(r_0,0)$ and $H_2=(-r_0,\pi/2)$,
which are hyperbolic (i.e., with one unstable and one stable
direction), $S=(-r_0,0)$, totally stable, and $U=(r_0,\pi/2)$, totally
unstable; here $r_0=\arccos f >0$.
The decomposition of the phase space into different dynamical regimes
depends on the value of $K_\perp$, which determines how much $\phi$
can depart from $0$ (see~\cite{non-monotonous_long_in-prep} for full
details):

%%%
%%% PHASE SPACE STRUCTURE
%%%

\paragraph{(i) Case of high $K_\perp$:}
all trajectories end at $S$, i.e., there are no limit cycles
($v=0$). This regime persists until $f=\fc$, at which point the pairs
$(H_1,S)$, $(H_2,U)$ merge and give rise to a saddle-node
bifurcation~\cite{guckenheimer_non-linear_bifurcations}.
For small $\delta\!f\equiv f-\fc>0$, we have $\alpha \partial_t r
\simeq \delta\!f+\frac 12 r^2$ in the region $r\simeq 0$ where the
trajectory spends most of its time, whence
$r(t)=\sqrt{2\delta\!f}\tan[t\sqrt{\delta\!f/2}]$ and one recovers the
standard depinning law with exponent $\beta=\frac 12$
(Fig.\ref{fig:homoclinic-scheme}.b).

\paragraph{(ii) Case of intermediate $K_\perp$:}
there is in addition a second critical force $\fc^\star<\fc$
(Fig.\ref{fig:homoclinic-scheme}.b).  For $f<\fc^\star$ all
trajectories end at $S$. For $\fc^\star<f<\fc$ one has bistability:
depending on the initial condition (Fig.\ref{fig:phase_space}),
trajectories either end at $S$ ($v=0$ branch in $v(f)$) or move along
an attracting limit cycle ($v>0$ branch).
In that case, the bifurcation is
homoclinic~\cite{guckenheimer_non-linear_bifurcations} and $\fc^\star$
is the value of the force for which the trajectory starting in an
unstable direction of $H_1$ returns exactly to $H_1$ (see
Fig.\ref{fig:homoclinic-scheme}.a).
The value of $\fc^\star$ depends on global aspects, in contrast to the
case~\emph{(i)}.
The depinning law of the $v>0$ branch is evaluated as follows
(see~\cite{non-monotonous_long_in-prep}): the trajectory spends most
of its time close to the hyperbolic point $H_1$, of positive Lyapunov
exponent $\Lambda$ so that $r(t)\simeq (f-\fc^\star)e^{\Lambda t}$.
The period $\tau$ thus verifies $2\pi/\kappa \sim
(f-\fc^\star)e^{\Lambda\tau}$ and up to factors:
\begin{equation}
 v\propto \big|\log (f-\fc^\star)\big|^{-1}
\end{equation}

% At this stage, one has understood the analogy between our system and
% the dynamics of a \emph{massive} damped particle in a tilted cosine
% potential (MDPP).  In that system, if the mass is low enough (our case
% \emph{(i)}) the motion is overdamped, while a larger mass (our case
% \emph{(ii)}) may allow the particle to cross barriers provided its
% momentum is high enough (and this depends on the initial condition,
% yielding bistability).
% %
% In fact the MDDP also presents a homoclinic
% bifurcation~\cite{guckenheimer_non-linear_bifurcations}.  In our
% problem, the phase $\phi$ thus plays the role of a momentum
% conjugated to the position, giving inertia to the particle.
% %
% However, the two problems cannot be identified, due to the
% presence of the mirror fixed points $S$ and $H_2$. 
% %~\cite{non-monotonous_long_in-prep}.
% %
% An important difference is also that a momentum is not periodic while
% $\phi$ is, which has dramatic consequences for the depinning
% law when $K_\perp$ takes lower values, as we shall now see.

%%% SUCCESSION OF BISTABLE TRANSITIONS

\begin{figure}[t] %%%% FIGURE NUMBER 4
  \centering
  \includegraphics[width=.48\columnwidth]{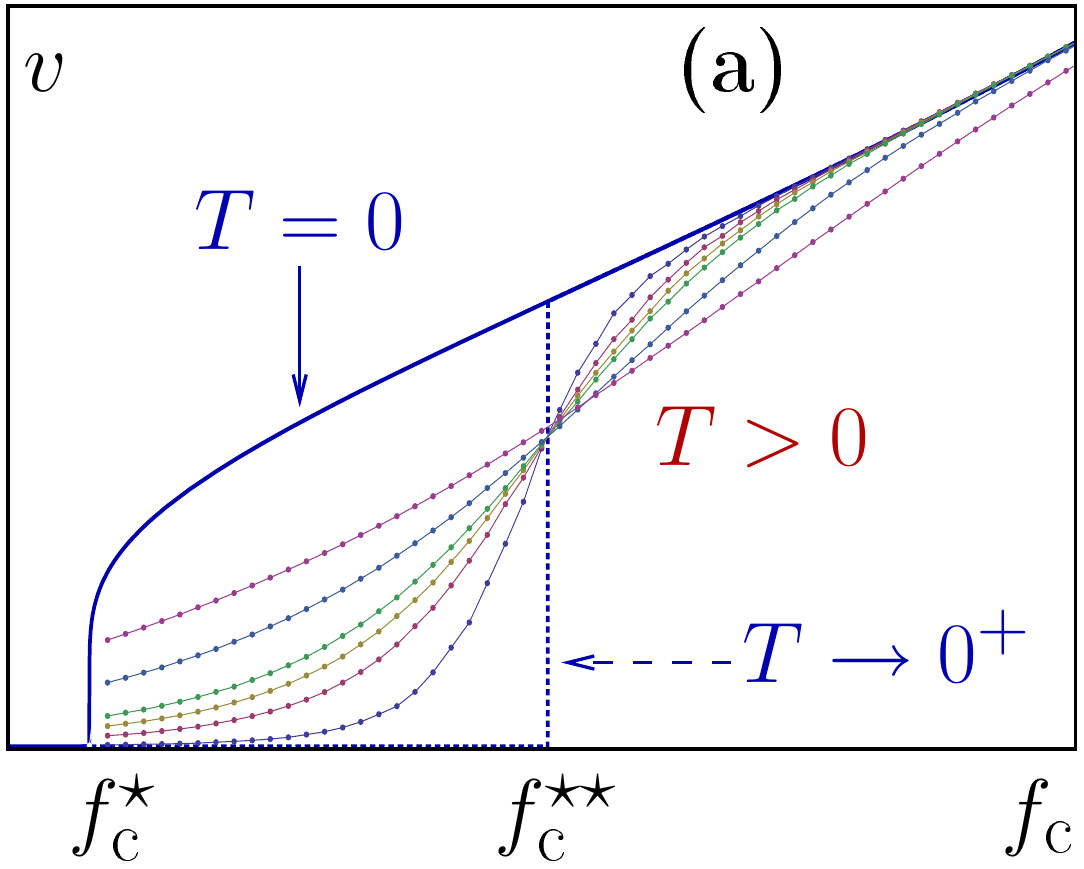}
  \includegraphics[width=.48\columnwidth]{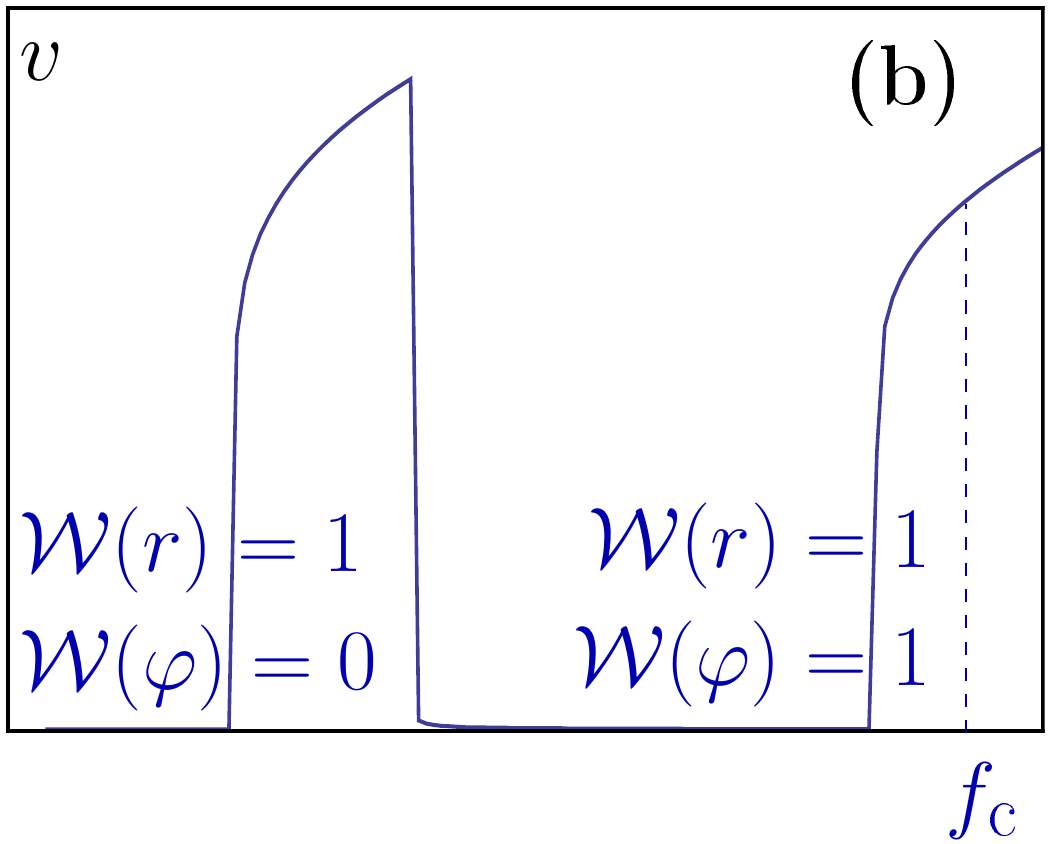}
  \vspace{-2mm}
  \caption{(color online) 
    \textbf{(a)} For any finite $T$ there exists a critical field 
    $\fc^{\star\star}$ such that  for $f<\fc^{\star\star}$ the albeit
    small probability of falling into the trap $S$ exceeds that of escape
     and the velocity is small while the inverse is true for $f>\fc^{\star\star}$.   
    The curves approach the vertical line as $T\to 0$ and all curves cross 
    at $f=\fc^{\star\star}$ to first order in $T$.
    \textbf{(b)}   For smaller $K_\perp$ (and $T=0$), successive bistable transitions
    occur, indexed by the winding numbers $\cW(r)$ and $\cW(\phi)$
    of $r$ and $\phi$ on one period. 
  }
  \vspace*{-3mm}
  \label{fig:depinn_T-dep}
\end{figure}

\paragraph{(iii) Case of smaller $K_\perp$:}
here, the particularly novel features appear, with a depinning law
$v(f)$ characterized by a succession of bistable regimes
(Fig.\ref{fig:depinn_T-dep}.b) separated by regions of zero velocity.
An interesting mechanism emerges: in general, the first bistable
regime is characterized by cycling trajectories with $r$ advancing and
$\phi$ oscillating within a bounded interval, but with increasing
force, $\phi$ will eventually rotate by a whole period of $\pi$, and
fall into $S$.
At this point there is a collision between the stable and unstable
limit cycles (of Fig.\ref{fig:phase_space}), and the original type of
limit cycle disappears for larger forces resulting in an intermediate
$v=0$ valley.
Increasing $f$ even more, the phase space reorganizes until there
appear trajectories with both $r$ and $\phi$ increasing with each
period of the limit cycle (Fig.\ref{fig:depinn_T-dep}.b).
Each bistable regime is governed by the same bifurcation as in case
$(ii)$, but is now also characterized by the number of windings of $r$
and $\phi$ during $\tau$.
This striking phenomenon, \emph{a topological transition}, arises from
the periodicity of the phase and cannot appear for instance in
situation where $r$ couples to an unbounded variable (e.g., the
momentum of a massive particle in a periodic potential).
But topological transitions can potentially be found in other systems,
e.g., for viscously coupled particles in a periodic
potential~\cite{ledoussal-marchetti_plastic-flow_prb2008}, described
by equations similar to~(\ref{eq:eq_motion1}-\ref{eq:eq_motion2}),
although $v(f)$ %was
is found to be monotonous for the conditions
of~\cite{ledoussal-marchetti_plastic-flow_prb2008}.

%%%
%%% TEMPERATURE AROUND THE HOMOCLINIC BIFURCATION
%%%

We finally address the finite temperature dynamics in the regime
$\fc^\star<f<\fc$, of particular interest since thermal fluctuations
cause the system to forget its initial condition, and thus destroy the
bistability.  Taking normal coordinates close to $H_1$,
(Fig.\ref{fig:homoclinic-scheme}) the evolution follows $\partial_t
\tilde r = \epsilon \tilde r + \tilde r^2 + \eta$,
$\partial_t\tilde\phi=-\tilde\phi$ 
(with $\epsilon\propto \fc-f>0$).  Starting from $H_1$ towards the
limit cycle and evolving with the noisy dynamics, the trajectory comes
back to $H_1$ with a Gaussian distribution of width~$\propto \sqrt{T}$
at distance $d(\epsilon)\propto\epsilon-\epsilon_c^\star$ from the
separatrix (Fig.\ref{fig:homoclinic-scheme}.a).
The mean escape time is determined by a competition between the large
Arrhenius time to escape from the local potential trap
$\tilde V(\tilde r)=-\epsilon {\tilde r}^2/2-{\tilde r}^3/3$
and the small probability $\sim\exp(-d(\epsilon)^2/T)$ of falling into
it~\cite{non-monotonous_long_in-prep}:
\begin{equation}
  \tau_{\text{escape}} \sim
  \exp\left(\frac{\epsilon^3}{3T}-A \frac{(\epsilon-\epsilon_c^\star)^2}{T} \right)
  \label{eq:escape-time_finiteT}
\end{equation}
Thus for $T>0$, the bistability curve is transformed in the following
manner: the curves $v(f,T)$ all cross at some new characteristic force
$\fc^{\star\star}$ (where the polynomial in $\epsilon$
in~\eqref{eq:escape-time_finiteT} has a zero)\footnote{This is at
  variance with the thermal rounding of the $v\sim(f-\fc)^\beta$ law,
  which develops no crossings for $T>0$.}.
For $f<\fc^{\star\star}$, the depinning is dominated by the escape
from the trap while for $f>\fc^{\star\star}$, $v(f,T)$ approaches the
positive branch of the $T=0$ law~(Fig.\ref{fig:depinn_T-dep}.a).
In the limit $T\to 0^+$, $v(f)$ is monostable and discontinuous in
$\fc^{\star\star}$, in contrast with the $T=0$ case.
Note that Vollmer and Risken~\cite{vollmer-risken_ZPB1980,risken_FP}
have studied the dynamics of a massive particle in a periodic
potential, for $T>0$, with results related to those of
Fig.\ref{fig:depinn_T-dep}.a,
but with an approach limited to that particular problem and only valid
in the small $\alpha$ regime.
In contrast, our approach not only displays a non-monotonous $v(f)$
but also allows for a general discussion of what happens in the
vicinity of a homoclinic bifurcation for any $\alpha$, in the
perspective of stochastic differential equations, and is more in the
spirit of Freidlin and
Wentzell~\cite{freidlin-wentzell_stochdiffeq,ETW_intermitt-noise_JPA1982}.

%%% LINK WITH EXPERIMENTS

Despite the oversimplified features of our model, it offers a possible
resolution of an experimental mystery~\cite{Parkin-SCI-2008}.
In the absence of pinning ($V=0$), and corresponding to the simple
Walker breakdown picture it is possible to reproduce the first peak of
$v(f)$ for parameters similar to those of~\cite{Parkin-SCI-2008}
($\alpha= 1.32\times 10^{-2}$, $K_\perp= 1500\,\text{Oe}$,
corresponding to $\fw=9.9\,\text{Oe}$, $\vw=200\,\text{m/s}$), but
$v(f)$ has only a peak without the valley seen in the experimental
data (Fig.\ref{fig:comparison_experiment}.b).
In contrast, due to the existence of a {\it topological transition},
for finite $V$ the velocity $v$ falls towards zero following a first
peak and only rises again for larger values the force $f$.
Indeed, simulations~\footnote{In simulations we discretize
  (\ref{eq:eq_motion1}-\ref{eq:eq_motion2}) in time ($>10^6$ steps
  with ${\text d}t=5\times 10^{-3}$) and average over 2048
  realizations. In both cases, the DW width was taken to
  $\lambda=15\,\text{nm}$~\cite{Parkin-SCI-2008} and the temperature
  to $T=300\,\text{K}$.  } of
(\ref{eq:eq_motion1}-\ref{eq:eq_motion2}) reproduce the `valley'
observed in $v(f)$ for similar parameters ($\alpha= 5\times 10^{-2}$,
$K_\perp= 1200\,\text{Oe}$, $V_0=50\,\text{Oe}$,
$\kappa=5.15/\lambda$).
We predict in particular the following watermark for the topological
transition: the second peak should coincide with the appearance of
non-zero $\langle\dot \phi\rangle $, measurable through the
emf~\cite{barnes_maekawa_prl2007} $\frac{\hbar}{e}\langle\dot
\phi\rangle$ (Fig.\ref{fig:comparison_experiment}.a) while in the
Walker picture $\langle\dot \phi\rangle>0$ immediately for $f>\fw$.

%
%%% CONCLUSION
%

To summarize, we have shown how the coupling between the phase and the
position of a rigid wall in 1D dramatically affects the depinning law,
which displays bistabilities and an unusual scaling $v\propto 1/|\log
(f-\fc^\star)|$.
Due to the periodicity of the phase, there are conditions for which
different bistable regimes follow one another with increasing $f$,
yielding for $T>0$ a \emph{non-monotonous} $v(f)$, which might well
explain features of recent measurements~\cite{Parkin-SCI-2008}.
It would be valuable to consider the current-driven case ($\vs\neq0$),
of interest in the context of spintronics.
Moreover, the solitonic ansatz can also describe interfaces with
non-zero dimension, where the interplay between the phase and the
elastic deformations potentially affects the creep motion and the
depinning.
%

%{\small [Also: 1/log law strangely reminiscent of a phenomenological fit by Kardar~\cite[Fig. 6]{kardar_phys-rep1998}?]}

%%% ACK

\begin{acknowledgments}
  This work was supported in part by the Swiss NSF under MaNEP and
  Division II.
\end{acknowledgments}

%\bibliographystyle{apsrev2}
%\bibliography{totphys,non-monotonous_depinn}

\end{document}